\documentclass[twocolumn,prd]{revtex4}%
\usepackage{graphicx}
\usepackage{amsmath}
\usepackage{amsfonts}
\usepackage{amssymb}%
\newcommand{\f}{\begin{equation}}
\newcommand{\ff}{\end{equation}}
\newcommand{\fa}{\begin{eqnarray}}
\newcommand{\ffa}{\end{eqnarray}}

\begin{document}
\title{Rainbow universe}
\author{Yi Ling$^{1,2}$}\email{yling@ncu.edu.cn}
\affiliation{${}^1$ Center for Gravity and Relativistic
Astrophysics, Nanchang University, 330047, China}
\affiliation{%
${}^2$ CCAST (World Laboratory), P.O. Box 8730, Beijing
   100080, China}

\begin{abstract}
The formalism of rainbow gravity is studied in a cosmological
setting. We consider the very early universe which is radiation
dominated. A novel treatment in our paper is to look for an
``averaged'' cosmological metric probed by radiation particles
themselves. Taking their cosmological evolution into account, we
derive the modified Friedmann-Robertson-Walker(FRW) equations which
is a generalization of the solution presented by Magueijo and
Smolin. Based on this phenomenological cosmological model we argue
that the spacetime curvature has an upper bound such that the
cosmological singularity is absent. These modified $FRW$ equations
can be treated as effective equations in the semi-classical
framework of quantum gravity and its analogy with the one recently
proposed in loop quantum cosmology is also discussed.

\end{abstract}

 \maketitle

\section{Introduction}

Any theory aiming to reconcile general relativity and quantum
mechanics is characterized by the presence of the Planck length
$(l_p=\sqrt{\hbar G/ c^3})$, which is the combination of the
gravitational constant, the Planck constant and the speed of light.
It has been long believed that this should be the minimum length
scale that one can observe or take measurement in laboratory. In a
cosmological setting it is also conjectured that the Planck length
would provide a bound to the curvature of spacetime such that the
big-bang singularity could be avoided in the framework of quantum
gravity.

In recent years the semi-classical or phenomenological effect of
quantum gravity has also been greatly
investigated\cite{Amelino00ge,Amelino00mn,Amelino03ex,Amelino03uc,Magueijo01cr,Magueijo02am,Smolin02sz,Smolin05cz}.
This is an essential step to link the fundamental quantum theory of
gravity which is under exploration and the classical general
relativity which however has proved to describe our low energy world
quite well. Particularly it is proposed that it may be responsible
for some puzzles occurring in recent astronomical and cosmological
observations, such as the threshold anomalies of ultra high energy
cosmic rays and Tev
photons\cite{Colladay98fq,Coleman98ti,Amelino00zs,Jacobson01tu,Myers03fd,Jacobson03bn,Amelino05ik}.
Then in this direction a key issue arises, namely how to construct a
semi-classical or effective theory of quantum gravity where the
Planck length may play a fundamental role but without violating the
principle of relativity. Among several candidates, doubly special
relativity is supposed to be such a framework to implement this
scheme. It is a deformed version of conventional special relativity
with two universal constants which are the same for all inertial
observers, one being the speed of light while the other the Planck
length or Planck energy. This can be accomplished by a non-linear
Lorentz transformation in momentum space, which leads to a deformed
Lorentz symmetry such that the usual energy-momentum relations or
dispersion relations in special relativity may be modified with
corrections in the order of Planck length. Nevertheless, in the
framework of doubly special relativity the definition of the dual
position space is not trivial due to the nonlinearity of the Lorentz
transformation. To overcome this difficulty, Magueijo and Smolin
proposed in\cite{Magueijo02xx} that maybe there is no single
spacetime background when the effect of moving probes on geometry
 is taken into account. Instead, it is replaced by a one
parameter family of metrics which depends on the energy of these
test particles, forming a ``rainbow'' metric. More explicitly,
corresponding to each modified dispersion relation, a dual rainbow
metric can be constructed. Consequently when incorporating curved
spacetime the connection and curvature are energy dependent such
that the usual Einstein's equations is modified as a one parameter
family of equations. This proposal has received a lot of attention
recently and other stimulated work on this formalism can be found in
\cite{Galan04st,Galan05ju,Hackett05mb,Aloisio05qt,Ling05bp,Galan06by,LHZ}.

As specific examples the modified version of $FRW$ solution and
Schwarzschild solution to these equations have been presented in
\cite{Magueijo02xx}. These solutions are derived in a general sense
that the probe can be any test particle moving in the spacetime.
Thus during the process of solving deformed Einstein's equations,
those correction terms in a rainbow metric, though dependent on the
energy of probes $\epsilon$, can be treated as a constant
independent of spacetime coordinates. In this paper we intend to
take the formalism of rainbow gravity as a semi-classical framework
of quantum gravity, and study its impact on the evolution of the
early universe. Rather than taking any specific measurement into
account, we intend to consider the semi-classical effect of
radiation particles on the spacetime metric during a longtime
process. Then the probe energy appearing in the correction term of
the rainbow metric is identified with the one of photons or other
sorts of massless particles like gravitons which dominate the very
early universe. In this setting the energy of probes obviously
varies with the cosmological time. In this paper we take this effect
into account and derive a generalized formalism of the modified
$FRW$ equations previously presented in \cite{Magueijo02xx}. Based
on this, we further propose a strategy to define an intrinsic metric
for the universe by considering the ``average'' effect of all
radiation particles. This phenomenological model provides the
spacetime curvature an upper bound such that the cosmological
singularity is absent in this semi-classical framework. We will end
this paper by pointing out the analogy of the modified $FRW$
equations presented here with the one recently proposed in loop
quantum cosmology.

\section{Review on Rainbow Gravity }
We start this section with a quick review on rainbow gravity
formalism proposed by Magueijo and Smolin\cite{Magueijo02xx}. In
deformed or doubly special relativity, the invariant of energy and
momentum in general may be realized by a non-linear Lorentz
transformation in momentum space, leading to a modified
energy-momentum relation or dispersion relations as  \f
\epsilon^2f_1^2(\epsilon,\eta)-p^2f_2^2(\epsilon,\eta)=m_0^2,\ff
where $f_1$ and $f_2$ are two functions of energy depending on the
specific formulation of boost generator and $\eta$ is a
dimensionless parameter. Now to incorporate the effect of gravity,
they proposed a deformed equivalence principle of general
relativity, stating that the free falling observers who make
measurements with energy $\epsilon$ will observe the same laws of
physics as in modified special relativity. Furthermore, the
correspondence principle requires that $f_1$ and $f_2$ approach to
unit as $\epsilon/E_{p}\ll 1$. On the other hand, the position space
is defined by requiring that the contraction between momenta and
infinitesimal displacement be a linear invariant. \f
dx^{\mu}p_{\mu}=dt\epsilon+dx^ip_i.\ff As a result, the dual space
is endowed with an energy dependent invariant \f ds^2=-{1\over
f_1^2(\epsilon,\eta)}dt^2+{1\over f_2^2(\epsilon,\eta)}dx^2.\ff Keep
going on and taking the effect of gravity into account, a one
parameter family of connections as well as curvature tensors can be
constructed from the rainbow metrics, leading to a modified
Einstein's field equations \f G_{\mu\nu}(\epsilon)=8\pi
G(\epsilon)T_{\mu\nu}(\epsilon)+g_{\mu\nu}(\epsilon)\Lambda
(\epsilon),\label{MEE}\ff where Newton's constant as well as the
cosmological constant is conjectured to be energy dependent in
general as one expects from the viewpoint of renormalization group
theory, for instance $G(\epsilon)=g^2(\epsilon)G$ and $\Lambda
(\epsilon)=h^2(\epsilon)\Lambda$.
\section{Modified FRW universe }
In this section we derive cosmological solutions to modified
Einstein equations (\ref{MEE}).  The modified FRW metric for a
homogeneous and isotropic universe may be expressed in terms of
energy independent coordinates as \f ds^2=-{1\over
f_1^2(\epsilon,\eta)}dt^2+{1\over
f_2^2(\epsilon,\eta)}a(t)\gamma_{ij}dx^idx^j,\ff where $\gamma_{ij}$
is the spatial metric. For simplicity we consider the spatially flat
universe with \f K=0\ \ \ \Lambda=0 \ \ \ \gamma_{ij}=\delta_{ij},
\ff but it is expected to be straightforward extending our analysis
to the other cases.

Usually one is free to pick up arbitrary particle  as a probe and
for any specific operation of measurement its energy $\epsilon$ can
be treated as a constant which is independent of spacetime
coordinates. However, in early universe of course no such real
measurement is taken. Here we intend to apply this mechanism to
study the semi-classical effect of radiation particles on the
background metric during its evolution which is a longtime process.
Then it is natural to take one of radiation particles as the probe.
Moreover, in this circumstance the evolution of energy $\epsilon$
with the cosmological time need to be considered, maybe denoted as
$\epsilon(t)$. Now taking this into account we derive the modified
$FRW$ equations following the standard process in general
relativity. For convenience we further take the following ansatz \f
\eta=1, \ \ \ f_2(\epsilon,\eta)=g^2(\epsilon)=1,\ff such that the
non-vanishing components of the associated connection read as \f
\Gamma^0_{00} = -{\dot{f}\over f}, \ \
      \Gamma^0_{ij} = f^2\dot{a}a\delta_{ij},\ \ \ \
      \Gamma^i_{0j} = \delta^i_j{\dot{a}\over a}. \label{ks}\ff
Going ahead we obtain the non-trivial components of Riemann tensor
as \fa
     R^i_{00j} &=& {\ddot{a}\over a}\delta^i_j+{\dot{a}\dot{f}\over af}\delta^i_j\nonumber\\
     R^0_{i0j} &=& f^2a\ddot{a}\delta_{ij}+f\dot{f}a\dot{a}\delta_{ij}\nonumber\\
     R^i_{jkm} &=& f^2\dot{a}^2(\delta^i_k\delta_{jm}-\delta^i_m\delta_{jk}).\ffa
Thus the Ricci tensor components are\fa
     R_{00} &=& 3{\ddot{a}\over a}+3{\dot{a}\dot{f}\over af}\nonumber\\
     R_{ij} &=& \left[ f^2\left( a\ddot{a}+2\dot{a}^2\right)+f\dot{f}a\dot{a}\right]\delta_{ij}\nonumber\\
     R &=& -6f^2\left( {\ddot{a}\over a}+{\dot{a}\dot{f}\over af}+{\dot{a}^2\over a^2}\right). \ffa
Next we consider the fluid with the energy-momentum tensor \f
T_{\mu\nu}=\rho u_{\mu}u_{\nu}+P(g_{\mu\nu}+u_{\mu}u_{\nu} )\ff
where $\rho$ denotes the energy density while $P$ the pressure, and
$u_{\mu}$ is defined as \f u_{\mu}=(f^{-1},0,0,0) \ff such that it
is a unit vector, namely \f u_{\mu}u_{\nu}g^{\mu\nu}=-1. \ff
 Applying this setting to the conservation equation of energy-momentum tensor \f
\nabla^{\mu}T_{\mu\nu}=0,\ff we find that \f
\dot{\rho}=-3H(\rho+P), \label{ce}\ff which is the same as the usual
one. It is also interesting to notice that this form is universal
for all comoving observers, while the connection or the covariant
derivative $\nabla(\epsilon)^{\mu}$ is energy dependent and thus the
symbols in (\ref{ks}) have been employed when deriving the equation
(\ref{ce}).

Now we turn to the modified Einstein equations. Substituting all the
terms obtained above into (\ref{MEE}) gives rise to the modified FRW
equations as \fa H^2 &=& {8\pi G\over 3}{\rho\over
 f^2}\label{mf}\nonumber\\
\dot{H} &=& {-4\pi G}{(\rho+P)\over
 f^2}-H{\dot{f}\over f},
\label{mf2} \ffa where we have introduced the Hubble factor
$H=\dot{a}/ a$. As in standard cosmology, from above equations we
may derive the conservation equation (\ref{ce}), which means only
two of these three equations are independent.

At the end of this section we point out that it is straightforward
to derive a general version of modified $FRW$ equations when both
$f_1$ and $f_2$ are general functions varying with the cosmological
time.
\section{Implications to the cosmological singularity}
In this section we consider the evolution of modified FRW universe.
For explicitness we take the function $f$ as \f
f^2=1-l_p^2\epsilon^2,\label{mfd}\ff which has also been considered
in the context of black hole physics in\cite{LHL,Ling05bp}. One
extraordinary property of this function is giving rise to a relation
\f \epsilon^2=\frac{1}{2l_p^2}\left( 1-\sqrt{1-4l_p^2p^2}\right),\ff
such that both the momentum and energy are bounded as $p\leq {1\over
2l_p} $,$\epsilon\leq {1\over \sqrt{2}l_p}$. As we stated above,
rather than picking up any specific particle from the radiation at
random, we more intend to take the radiation as an ensemble and
consider the ``average'' effect of radiation particles which
dominate the very early universe. Such a scheme has been applied to
investigate the thermodynamics of modified black holes as well. For
details we refer to \cite{Ling05bp,Galan06by} and the previous
relevant work in \cite{Adler01vs}. Then the energy appearing in
rainbow metric is identified with the statistical mean value of all
massless radiation particles. As a result, the modified dispersion
relation varies with the cosmological time \f
f^2=1-l_p^2\bar{\epsilon}^2(t).\ff Then the conservation equation
(\ref{ce}) and the $FRW$ equation (\ref{mf}) read as \f
{\dot{\rho}\over \rho}=-3(1+\omega){\dot{a}\over a}, \label{frw2}
\ff \f {\dot{a}^2\over a^2}={8\pi G\over 3}{\rho\over
 1-l_p^2\bar{\epsilon}^2}.\ff
Before explicitly describing the evolution of the universe subject
to equations (\ref{frw2}), we still need to know the equations of
state of the radiation. As pointed out in
references\cite{Magueijo02xx,Alexander01ck}, the usual relation of
$\rho$ and $P$ would be modified due to the modified dispersion
relations, namely $\omega\neq 1/3$ exactly. However, as we show in
the appendix, when the temperature is below the Planck scale this
modification will only provide correction terms in higher orders of
the Planck length, which will not change the main picture we
illustrate below. Furthermore, from the dynamical point of view we
find in this semi-classical framework, the temperature of the
universe will also be bounded in the very early epoch, which is
about the Planck scale such that the case of $T\gg 1/l_p$, though
discussed in our appendix, will not occur in the evolution history
of the universe. In hence for our purpose, it is still plausible to
identify the energy of radiation particle with the density through
the conventional relations \f \bar{\epsilon}^2\sim \sqrt{\rho} \ \ \
\ \ {\dot{\rho}\over \rho}\sim -4{\dot{a}\over a}.\label{ai}\ff Thus
we have an equation \f {l_p^2\sqrt{1-x}\over x^2}\dot{x}=C\ff where
$x=l_p^2\sqrt{\rho}$ and $C$ is a constant . The solution is \f
t={l_p^2\over C} \left[ {\sqrt{1-x}\over x}+\ln {(1-\sqrt{1-x})\over
\sqrt{x}} \right]\label{eod}\ff At the classical limit, namely
$l_p\rightarrow 0$, it is easy to find that the expression goes back
to the usual one $\rho\sim {1\over t^2}$, but in early time of the
universe $t\rightarrow 0$, we find the density will go to a finite
number rather than a divergent disaster \f \rho\sim \left( {l_p\over
l_p^2+Ct}\right)^4\sim {1\over l_p^4}.\ff Thus in this effective
theory the cosmological singularity is absent. We stress that the
avoidance of singularity in this context is not apparent. Time
reparametrization will not change this conclusion and this can be
seen from the equation (\ref{eod}).

Of course above consideration based on the identification in
(\ref{ai}) is the zeroth approximation and quite sketchy. More
quantitative illustration of the evolution may be obtained through
numerical simulation based on equations $(\ref{A1})-(\ref{A4})$ in
the appendix.
\section{Discussions and conclusions}
It must be pointed out that the complete understanding on the fate
the cosmological singularity calls for a complete quantum theory of
gravity. Our analysis presented here, though in a heuristic manner,
has shed light on this issue and reflected the important role played
by non-perturbative gravity around the Planck scale. One of our
motivations discussing the fate of the cosmological singularity in
semi-classical regime also comes from recent progress made in loop
quantum cosmology, where the exact solutions to quantum equations
are remarkably well approximated by a naive extrapolation of some
semi-classical effective equations. It is interesting to compare the
effective $FWR$ equations from rainbow gravity with those from loop
quantum cosmology, where the effective equations have the form
\cite{Ashtekar06rx,Ashtekar06uz,Singh06im}\f H^2={8\pi G\over 3}\rho
\left( 1-{\rho\over \rho_c}\right)\label{lqc1}\ff \f \dot{H} =
{-4\pi G}(\rho+P) \left( 1-2{\rho\over \rho_c}\right) \label{lqc2}
\ff and $\rho_c\sim l_p^{-4}$. They may be obtained from the eqs.
(\ref{mf}) and (\ref{mf2}) if we naively require that \f
f^2=(1-l_p^4\epsilon^4)^{-1}.\ff Therefore, our scheme presented
here indicates a potential relation between the rainbow gravity
formalism and the effective theory of loop quantum gravity at the
semi-classical level \footnote{It is worthwhile to point out that
the naive identification involves more subtleties to clarify from
the side of doubly special relativity. First the corresponding
dispersion relation which has a higher order correction term does
not manifestly provide an upper bound on either the momentum or
energy of a single particle. Second the resulting density of states
from (\ref{dos}) diverges as the energy approaches to the Planck
scale, implying some certain cutoff of the energy should be put in
by hand which sounds not satisfying. }.

In summary, we have studied the semi-classical effect of radiation
particles on the background metric in the framework of rainbow
gravity, and obtained the modified $FRW$ equations where the energy
of particles varies with the cosmological time. When taking some
specific modified dispersion relation into account, we find the
evolution picture of the universe in the early era dramatically
changes and the cosmological singularity can be avoided. The scheme
presented here may also be extended to more general cases, for
instance the spatial metric is non-flat and the cosmological
constant can also be involved. In particular it is supposed that the
scalar field should play an important role in very early universe,
its effective theory in the formalism of rainbow gravity as well as
the effect of the energy dependence of the Newton's constant is
under investigation.

\section*{Acknowledgement}
 I am grateful to Song He, Bo Hu, Xiang Li, Xiao-qing Wen and Hongbao Zhang for a lot of
 discussions and helpful suggestions. This work is partly supported by NSFC (No.10405027) and SRF for ROCS, SEM.

\section{Appendix: Statistics for radiation particles with modified dispersion relations }
The thermodynamical feature of radiation particles with modified
dispersion relations has been considered previously in the context
of non-commutative geometry\cite{Alexander01ck}. Here we start with
the general form of modified dispersion relations as considered
above \f f^2\epsilon^2-p^2c^2 =0. \ff Thus we have \f dp={f\over
c}\left( 1+\epsilon{f'\over f} \right) d\epsilon. \ff Then the usual
density of states is modified as  \fa
g(\epsilon)d\epsilon &=& 2{V\over h^3}4\pi p^2dp \nonumber\\
&=& {8\pi V\over h^3c^3}f^3\left( 1+\epsilon{f'\over f}
\right)\epsilon^2 d\epsilon. \label{dos}\ffa Consequently the
following statistical quantities are modified as \f
\bar{N}=\int{g(\epsilon)\over
e^{\epsilon/kT}-1}d\epsilon.\label{nop}\ff
 \f \rho={\bar{U}\over V}=\int{\epsilon\over V}{g(\epsilon)\over
e^{\epsilon/kT}-1}d\epsilon.\ff
 \f {PV\over kT}=\int -\ln (1-e^{-\epsilon/kT})g(\epsilon)d\epsilon.\ff
 Plugging the modified density of states into the equation (\ref{nop}) we obtain
 \fa \bar{N}&=& \int {8\pi V\over h^3c^3}f^3\left(
1+\epsilon{f'\over f} \right){\epsilon^2\over
e^{\epsilon/kT}-1}d\epsilon \nonumber\\
&=&{8\pi V\over h^3c^3}\int_0^{1\over \sqrt{2}l_p}
{\epsilon^2(1-l_p^2\epsilon^2)^{1\over 2}(1-2l_p^2\epsilon^2)\over
e^{\epsilon/kT}-1}d\epsilon. \ffa Now we consider above integrations
with the specific function given by (\ref{mfd}). For low
temperature, we may redefine $ x\equiv {\epsilon\over kT}$ and
$\lambda \equiv l_pkT\ll 1$, such that the integral may be
approximately evaluated as \f \bar{N}\propto VT^3(1-\alpha
l_p^2T^2+...).\label{A1}\ff Similarly we can get the results for the
density and the pressure
 \f \rho \propto \sigma T^4(1-\beta l_p^2T^2+...),\label{A2}\ff
\f P \propto {1\over 3}\rho(1-\gamma l_p^2T^2+...),\label{A3}\ff
providing modifications to both Stefan-Boltzmann law and the
equation of state, where the parameters may approximately evaluated
as $\alpha=25.88$, $\beta=46.95$ and $\gamma=28.17$. As a result, we
notice that \f \bar{\epsilon}={\rho V\over \bar{N}}\propto
kT(1+\delta l_p^2T^2+...).\label{A4}\ff For high temperature
$\lambda\gg 1$, however, these relations will change greatly. It can
be found that \f \bar{N}\propto {8\pi V\over h^3c^3}{kT\over
l_p^2}\int_0^{1\over \sqrt{2}} 2x(1-x^2)^{1\over 2}(1-2x^2)dx, \ff
and \f \bar{U}\propto {8\pi V\over h^3c^3}{kT\over
l_p^3}\int_0^{1\over \sqrt{2}} 3x^2(1-x^2)^{1\over 2}(1-2x^2)dx, \ff
such that the energy of particles is bounded as we expect from the
viewpoint of doubly special relativity\f \bar{\epsilon}\sim
{\bar{U}\over\bar{N}}\sim {1\over l_p},\ff which is independent of
the temperature.

\end{document}